\documentclass[runningheads]{llncs}
\usepackage{float}
\usepackage[dvips]{graphicx}
\usepackage{array}
\usepackage{multirow}
\usepackage{alltt}
\usepackage{float}
\usepackage{color}  
\usepackage{listings} 
\usepackage{paralist}  
\usepackage{acronym}
\usepackage{moreverb}
\usepackage{sidecap}
\usepackage{amsmath,amsfonts,amssymb}
\usepackage{calc}
\usepackage{ifthen}
\usepackage{textcase}
\usepackage{textcomp}
\usepackage{url}
\usepackage{xspace}

\usepackage{macroSO}
\usepackage{macrogentra2}

\begin{document}


\mainmatter 
\title{Generic Traces and Constraints, \\ GenTra4CP revisited}

\titlerunning{Constraints and Traces}

\author{Pierre Deransart}

\authorrunning{Deransart P.}

\institute{{\sc Inria} Rocquencourt, BP 105, 78153 Le Chesnay Cedex, France\\
\email{Pierre.Deransart@inria.fr}
}

\maketitle

\vspace{2mm}
\begin{abstract}
The generic trace format GenTra4CP has been defined in 2004 with the goal of becoming a standard trace format for the observation of constraint solvers over finite domains. It has not been used since.
This paper defines the concept of generic trace formally, based on simple transformations of traces.
It then analyzes, and occasionally corrects, shortcomings of the proposed initial format and shows the interest that a generic tracer may bring to develop portable applications or to standardization efforts, in particular in the field of constraints.
\end{abstract}

\section{Introduction}
\label{intro}

Following the RNTL OADymPPaC project \cite{oadymppac}, a generic trace format, called GenTra4CP (Generic Trace for CP), has been proposed in 2004 in order to specify traces of CSP(FD) resolution. One of the objective was to allow the development of portable powerful tools for solvers analysis. 
This format was designed as a kind of standard, consisting of a precise syntax of trace events including an XML DTD, and an operational semantics, called observational semantics, which is a partial operational semantics applicable to a set of finite domains solvers.

Such ``standard'' conforming tracers were implemented in four solvers. Several tools for analysis of resolution and search strategies were developed in four different environments, just using the generic trace GenTra4CP. They have been used with success after a minimal customization work for each of them. 
However, at that time, no formal characterization of the generic nature of the trace format has been given. 
Even if the implementation of the tools starting from a well defined generic trace could be realized without difficulty, and even if there was obtained a considerable gain in portability, it was virtually impossible to assess in advance the effort of adaptation needed for a solver to use the tools. Moreover it was not always possible to figure out exactly what some tool was actually observing. GenTra4CP format has been used only in the project in which it has been defined.

This paper attempts to overcome these limits, by defining formally the concept of generic trace.
It analyzes formally the nature of the generic format GenTra4CP and its limitations that could have made difficult its broader use. It also shows the interest that a generic tracer approach may bring for standardization efforts and portability of applications, in particular for constraints, in proposing an approach of trace based semantics grounded on a partial operational semantics.

After an introductory section on operational semantics of traces, the Section~\ref{sec:stabstr} introduces some simple relations between traces in order to formalize the concept of generic trace, and to provide a proof method of compliance of a particular process trace with the generic one. The Section~\ref{sec:tragen} explains the generic approach and its interest for portability. The Section~\ref{sogentra4cp}  applies this approach to the case of GenTra4CP verifying the compliance of a particular solver. The formal description of the GenTra4CP trace format is borrowed from \cite{oadymppac} and from \cite{ercimlnai} for the solver. This allows a better understanding of the strengths and limitations of this approach as introduced in 2004. We can then establish a possible link between the efforts of constraints standardization, and the specification method based on generic trace. 


\section{Preliminaries}
\label{sec:prelim}

A {\em trace} object consists of an initial state $s_0$ followed by an ordered finite or infinite sequence of {\em trace events}, denoted $<s_0, \overline{e}>$. ${\cal T}$ is a set of traces. A {\em prefix} (finite, of size $t$)  of a trace $T~=~<~s_0, \overline{e_n}>$ (finite or infinite, here of size $n \geq t$) is a partial trace $U_t = <s_0, \overline{e_t}>$ which corresponds to the $t$ first events of $T$, with an initial state at the beginning. A prefix consisting of just an initial state is of size 0. The set of all the prefixes of ${\cal T}$ is denoted $Pref({\cal T})$, with ${\cal T} \subseteq Pref({\cal T})$.

Every trace can be decomposed into segments containing trace events only, except prefixes which start with a state. An associative operator of concatenation may be used to denote sequences concatenations. The neutral element is $\epsilon$ (empty sequence). A trace may describe state transitions, such that a segment may also be represented as $s T_t s_t$ where $s$ is the state in which the sequence $T_t$ starts and  $s_t$ the state reached after the last trace event in the sequence. A segment (or prefix) of size 0 is either an empty sequence or a state.

A {\em domain of traces} over ${\cal T}$, ${\cal DT}_{\cal T}$, is a set whose elements are sets of all prefixes of one or more traces of ${\cal T}$. An element is prefix closed. Such a set is closed by union and intersection, and, two included elements are such that the smaller contains all the prefixes of some traces of the largest. A trace domain is a complete lattice denoted ${\cal DT_T}(\subseteq,\bot,\top,\cup,\cap)$ where $\bot$ is the empty set and $\top = Pref({\cal T})$.

Traces are used to represent the evolution of systems by describing the evolution of their state. We will distinguish two kinds of traces:
\begin{itemize}
\item the {\em virtual traces} (${\cal T}^v$) whose events have the form $e = (r, s)$ where $r$ is a {\em type of action} associated with a state transition and $s$, called {\em virtual state}, the new state reached by the transition and described by a set of {\em parameters}. Virtual trace corresponds to sequences of states of an observed system. 
\item the {\em actual traces} (${\cal T}^w$) whose events have the form $e = (a)$ where $a$ is an {\em actual state} described by a set of {\em attributes}. Actual traces corresponds to sequences of events produced by a tracer of an observed system. Thy usually encode states changes in a synthetic manner. 
\end{itemize}


We give here a simplified but sufficient definition of observational semantics. More general definitions can be found in \cite{TMTmanuscript11e}.

\begin{definition} [Observational Semantics]
\label{def:MI}

An observational semantics consists of $< S, R, A, T, E, I, S_0 >$, where

\vspace{1mm}
\begin{itemize}
\item $S$: {\em domain of virtual states},
\item $R$: finite set of {\em action types}, set of identifiers labeling the transitions.
\item $A$: {\em domain of actual states},
\item $T$: {\em state transition function} $T : R$ {\tt x} $S \rightarrow S$, denoted $T(r, s) = s'$ or  $T(r, s, s')$ if it is a relation,
\item $E_l$: {\em local trace extraction function} $E_l: S \times R  \times S \rightarrow A$,
\item $I_l$: {\em local trace reconstruction function} $I_l: S \times A \rightarrow R \times S$,
\item $S_0 \subseteq S$, set of {\em initial states}.
\end{itemize}
\end{definition}

The extraction and reconstruction functions can be extended into functions $E$ (resp. $I$) between sets of virtual and actual traces, and must verify the relation of {\em faithfulness}, $I = E^{-1}$. 
Local and extended functions satisfy the properties:

$E(s_0 e_1 ... e_i ...) = s_0 E_l(s_0, r_1, s_1) ... E_l(s_{i-1}, r_i, s_i) ...$ with $E_l(s_{i-1}, r_i, s_i) = a_i$, and

$I(s_0 a_1 ... a_i ...) = s_0 I_l(s_0, a1) ... I_l(s_i, a_{i+1}) ...$ with $I_l(s_{i-1}, a_i) = (r_i, s_i)$.

\vspace{1mm}
The local and transition functions may be represented by rules as illustrated by the Figure~\ref{fig:reduceex}. 

\vspace{1mm} 
The observational semantics of an observed process can be considered as an abstraction of some refined operational semantics \cite{CousotPOPL02}. This relation will be expressed here as a relation between domains of traces.
Such a relation may be expressed either between virtual or actual traces. 
Due to the faithfulness property, the abstraction function $D_w$ on actual traces verifies with $D_v$, the abstraction function on virtual traces, the following relations: $D_v = E_c \circ D_w \circ I_d$ and $D_w = I_c \circ D_v \circ E_d$. 

In the following it will be assumed that the faithfulness property is satisfied, whatever is the abstraction level of the trace description. In this case, the extraction function is deducible from the reconstruction one and reciprocally. Therefore it is sufficient to specify the transition function with the extraction only or with the reconstruction. 
\begin{figure}[ht]
\footnotesize 
\noindent
\def\et{,\ \ }
\begin{center}
\reglecontroleshortdx{\reduce{}}
{< {\cal D}(v) \et S_e \et A> }
{<{\cal D}(v)-\Delta_v^c \et S_e \cup \bar{a} \et A'>}
{\left\{\begin{array}{l}
\hbox{remove}\ \Delta_v^c) \et  
a\ \hbox{wake up}\ c \\
(c, a) \in A \et 
A'=A-\{(c, a)\} \\
v \in \Var{c} \et 
\hbox{generate}\ \bar{a}
\end{array} \right\}}
\saut
\reglecontroleshortdx{\reduce{}}
{
< {\cal D}(v) \et S_e \et A' \cup \{(c, a)\} > \ \rightarrow\ 
<  {\cal D}'(v) \et S'_e \et A' >
}
{[ \reduce{} \et c \et v \et (S'_e - S_e) \et ({\cal D}(v) - {\cal D}'(v)) \et a]}
{\left\{\begin{array}{l}
\end{array} \right\}}
\saut
\reglecontroleshortdx{\reduce{}}
{[ \reduce{} \et c \et v \et \bar{a} \et \Delta_v^c \et a]}
{
< {\cal D}(v) \et S_e \et A > \ \rightarrow\ 
<  {\cal D}(v) - \Delta_v^c \et S_e \cup \bar{a} \et A - (c, a) >
}
{\left\{\begin{array}{l}
\end{array} \right\}}
\end{center}
\caption{Example of description of \reduce{} in the OS of GenTra4CP (Section~\ref{sogentra4cp}) with transition rule, extraction and reconstruction. Computations are specified on the right side}
\label{fig:reduceex}
\end{figure}
In practice, only actual traces are manipulated by the users, but thanks to the faithfulness property, for validation purposes, the virtual trace may be used.


\section{Abstraction relations: subtraces and derivations}
\label{sec:stabstr}

We introduce simple transformations on traces: subtraces and derivations. As it is sufficient to describe transformations on the virtual traces, they are described using one part of their description, namely $< S, R, T, S_0 >$ only.

Subtraces are obtained by considering a subset of parameters.

\begin{definition} [Subtrace]
\label{def:stacceptable}

Given a set of virtual traces  ${\cal T}$ defined by $< S, R, T, S_0 >$, 
if $S' \subseteq S$ is defined on a subset of parameters which do not depend\footnote{A parameter $p$ depends on $p'$ iff $p'$ is used in the computation of $p$ in some transition.} on any other parameter of $S-S'$, $R' \subseteq R$ is a subset of action types which use or modify these parameters only such that no other action type of $R-R'$ modifies them, $S'_0$ is the restriction of $S_0$ to $S'$, and $T'$ the restriction of $T$ to $S'$ and $R'$, then the set of traces ${\cal T}'$ defined by $< S', R', T', S'_0 >$ is a (parametric) subtrace of ${\cal T}$, denoted $Sub_P({\cal T}, {\cal T}')$.
\end{definition}
Note: it is possible that $S' \subseteq S$ and $R' = R$ ($S-S'$ contains redundant parameters, i.e. which depend only on the other parameters and thus may be removed).

\vspace{1mm}

\begin{definition} 

{\bf (Derivation field and derived trace)} 
\label{def:tracderivee}

Given two sets of traces ${\cal T}_c$ and ${\cal T}_d$, where ${\cal T}_c$ and ${\cal T}_d$ are said respectively {\em concrete} and {\em derived}, ${\cal T}_d$ is a derivation field of ${\cal T}_c$ by $D$ if there exists a mapping 
 $D: Pref({\cal T}_c) \rightarrow Pref({\cal T}_d)$, called a {\em derivation}, such that
for all finite derived prefixes $t_d$ of size $n$ and for all concrete prefix $t_c$ such that $D(t_c)=t_d$, there exists an increasing chain of concrete prefixes 
 $[t^0_c, t^1_c, ..., t^i_c, ..., t^{n-1}_c, t_c]$ (not necessarily contiguous), such that
\begin{itemize}
\item $D(t^0_c) \in S_{0,d}$, 
\item $\forall i > 0$ if  $D(t^i_c) = t^i_d$ with $t^i_d$ prefix of $t_d$ made of the $i$ first events, then $D(t^{i+1}_c) = t^{i+1}_d$.
\end{itemize}
If $D$ is surjective, 
the set ${\cal T}_d$ is called {\em derived trace} by $D$ of ${\cal T}_c$, noted  $Drv_D({\cal T}_c, {\cal T}_d)$.
\end{definition}

\vspace{1mm}
As defined, $D$ is a partial function. It can be made total by considering that all elements of $S_{0,c}$ have an image in $S_{0,d}$ and that the image of each prefix between $t^i_c$ and $t^{i+1}_c$ is $D(t^i_c)$.

\begin{figure}[h]
\begin{center}
\includegraphics[width=0.8\linewidth]{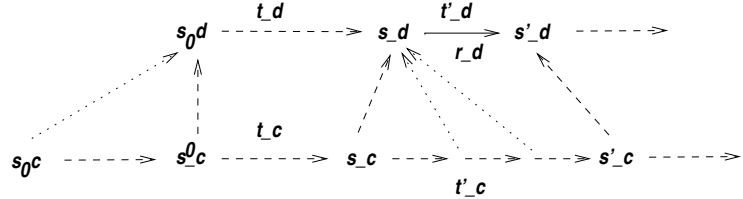}
\end{center}
\caption{Derivation (dashed arrow  correspond to the totalized derivation), $t_x$ denotes a prefix}
\label{fig:diagabstr}
\end{figure}

This approach puts emphasis on traces without considering the way they have been produced or the way they are specified. The main idea is that a trace transformation is the result of a computation on likely full prefixes of concrete traces, represented by de derivation $D$.



\begin{property}
\label{prop:deriv}
\ \ 

\vspace{1mm}
Given two derivations $D_1$ and $D_2$, if $D_1$ is surjective or if $D_2$ is total, $D_1 \circ D_2$ is a derivation.

A parametric subtrace (definition~\ref{def:stacceptable})  is a derived trace.
\end{property}

The following establishes a method of proof that two sets of traces specified by transition relations are related by a derivation.

\begin{definition} 

{\bf (Simulable Trace)}
\label{def:tracesimul}

Given two sets of traces ${\cal T}_c$ (concrete) and  ${\cal T}_d$ (derived), respectively defined with $< S_c, R_c, T_c, S_{0,c} >$ and $< S_d, R_d, T_d, S_{0,d} >$, ${\cal T}_c$ is {\em simulable} in ${\cal T}_d$ if $R_c$ and $R_d$ are in a one-one mapping $h$, and if there exists an application $d: S_c \rightarrow S_d$ such that:
\begin{itemize}
\item $\forall s_0 \in S_{0,c}, d(s_0) \in S_{0,d}$.
\item $\forall r_c \in R_c, s_c, s'_c \in S_c,\ T_c(s_c, r_c, s'_c) \Rightarrow \exists s_d, s'_d \in S_d,  d(s_c) = s_d \wedge d(s'_c) = s'_d \wedge T_d(s_d, h(r_c), s'_d)$.
\end{itemize}
\end{definition} 

\begin{theorem}
\label{prop:proofmethabstr}
\ \ 

Given two sets of traces ${\cal T}_c$ (concrete) and  ${\cal T}_d$ (derived), such that ${\cal T}_c$ is simulable in ${\cal T}_d$, then
${\cal T}_d$ is a derivation field for ${\cal T}_c$ and the corresponding derivation is total.
\end{theorem}

\begin{corollary}
\label{cor:proofmethabstr}
\ \ 

Given two sets of traces ${\cal T}$ and  ${\cal T}'$  such that there exists a parametric subtrace of ${\cal T}$ 
simulable in  ${\cal T}'$, then ${\cal T}'$ is a derivation field for ${\cal T}$.
\end{corollary}

\section{Generic Trace}
\label{sec:tragen}

The idea of generic trace meets the needs of specification and portability. It is intended to specify a process or an algorithm by its observable behavior, i.e. the trace of abstracted operations that it is expected to implement. The level of description must be general enough to include family of processes, and the level of granularity must be sufficiently refined to be used by a family of applications. This may be the case for example for applications such as monitoring, debugging, visualization tools, or any application using the generic trace. 

\begin{definition} [Generic Trace (GT)]
\label{def:trgen}

Given a family of processes $p \in P$, each of them equipped to produce traces ${\cal T}_p$, a set of traces ${\cal T}_g$ is generic if, for each process $p$ in the family, there exists a derivation $D_p$ of its traces which is a parametric subtrace of ${\cal T}_g$, that is:

$\forall p \in P, \ \exists\  {\cal T}$ such that 
 $Drv_{D_p}({\cal T}_p, {\cal T}) \wedge Sub_P({\cal T}_g, {\cal T})$.
\end{definition}

Three questions are then worth posing:
\begin{itemize}
\item How to ensure that the trace produced by some process is compliant with the GT?
\item Can the GT be used in application development, with the guarantee that the application will work with any compliant process?
\item Can the GT be extended to handle more processes in such a way that existing applications will still work?
\end{itemize}


Here are some possible answers.

\vspace{1mm}
\underline{Compliance to the Generic Trace}

A trace of a process is compliant w.r.t. the GT if it satisfies the definition~\ref{def:trgen}, i.e. there exists a subtrace of the GT which is a derivation of a subtrace of those of the process. It is thus possible either to implement straightforwardly the GT as it is (in this case the process produces exactly the GT), or to prove that the traces a process $p$ may generate verifies $\exists {\cal T}', Drv_{D_p}({\cal T}_p, {\cal T}') \wedge Sub_p({\cal T}_g, {\cal T}')$.

\vspace{1mm}
\underline{Building tools with the Generic Trace}

\begin{figure}[h]
\begin{center}
\includegraphics[width=0.6\linewidth]{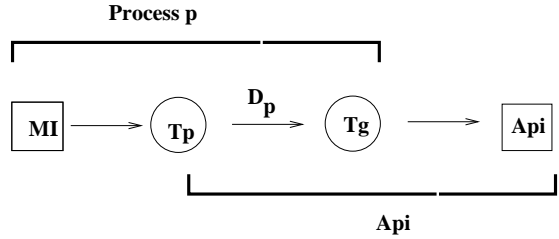}
\end{center}
\caption{Use of a Generic Trace: process or application adaptation}
\label{fig:utilgen}
\end{figure}

The interest of a generic trace is that it facilitates the development of tools that can be used with all compliant processes. The development is made considering that the tool uses at least a sub-GT covering sufficiently many processes. 
Thus it is possible to adapt the tool to the process $p$ by applying to the trace generated by the process (without any modification) the derivation $D_p$ to get a GT. This can be done at le level of the process (process can use any tool) or at the level of the tool (tool can be run with this particular process). The Figure~\ref{fig:utilgen} illustrates these two ways to adapt processes with compliant tracer and tools.

The fact that the GT has a formal specification makes it possible to realize a prototype (executable specification) which shall be itself a new compliant process. It is thus possible to use such a prototype to develop and test tools. This development method guarantees that any tool made on the top of the GT will be able to work with any compliant processes.

\vspace{1mm}
\underline{Generic Trace extensions}

As long as an extension of the GT preserves the fact that a process is compliant w.r.t. a subtrace of the extended GT, they still are compliant w.r.t. the extended GT. It is sufficient to ensure that any GT extension preserves the parametric subtraces. This guarantees that the compliant processes will continue to be usable by tools using the original GT.

%


\section{The Generic Trace GenTra4CP}
\label{sogentra4cp}

In the final document \cite{oadymppac}, the generic trace GenTra4CP is defined with an observations semantics whose transition function is defined with a subset of parameters. Thus only a generic subtrace has a formal semantics. The other parameters are described informally by the description of other attributes of the actual trace. Their syntax is fixed by a DTD XML and informal explanations are provided for each new attribute. We recall here the semantics as originally presented in \cite{oadymppac} (section 3.3.1)\footnote{Here one uses $n$ instead of $\nu$ to denote the current node of the choice-tree.}

\vspace{1mm}
{\tt Beginning of Citation}:
\begin{definition}(Solver State)

A solver  state is a 8-tuple:\quad
$\mathbb{S} = ({\cal V}, {\cal C}, {\cal D}, A, E, R, S_c, S_e)$\\
where:\quad
${\cal V}$ is the set of declared variables; \quad
${\cal C}$ is the set of declared constraints;\quad
${\cal D}$ is the function that assigns to each variable in ${\cal V}$
its domain (a set of values in the finite set $D$);\quad
$A$ is the set of active pairs of the form (constraint, solver
event\,\footnote{This work inherits from two areas, constraint solving
  and debugging, which both use the word ``event'' in correlated but
  different meanings: a {\em solver event} is produced by the solver
  and has to be propagated (e.g. the update of the domain bounds of a
  variable); a {\em trace event} corresponds to an execution step
  which is worth reporting about.});\quad $E$ is the set of solved
constraints;\quad $R$ is the set of unsatisfiable (or rejected)
constraints.
$S_c$ is the set of sleeping constraints;\quad
$S_e$ is the set of solver events to propagate (``sleeping
events'').
\end{definition}

$A$, $S_c$, $E$ and $R$ are used to describe four specific states of a
constraint during the propagation stage: active, sleeping, solved or
rejected.  

The {\em store of constraints} is the set of all
constraints taken into account. The store is called $\sigma$ in the
following and defined as the partition $\sigma=\{c \mid \exists (c,a)
\in A\} \cup S_c \cup E \cup R$. All the constraints in $\sigma$ are
defined, thus $\sigma \subseteq {\cal C}$.
The set of variables involved in the constraint $c$ is denoted by
$\Var{c}$. 
The predicate $false(c,{\cal D})$ (resp. $solved(c,{\cal D})$) holds
when the constraint $c$ is considered as unsatisfiable (resp.
solved: it is universally true and does not influence further reductions any more) by the domains in ${\cal D}$.

The search is often described as the construction of a search-tree.

\begin{definition}(Search-Tree State)

  The search-tree is formalized by a set of ordered labeled nodes ${\cal N}$ representing a tree,
  and a function $\Sigma$ which assigns to each node a solver state.
  The nodes in ${\cal N}$ are ordered by the construction.  Three
  kinds of nodes are defined and characterized by three predicates:
  failure leave ($failed({\mathbb S})$), solution leave
  ($solution({\mathbb S})$), and choice-point node 
  ($choice\hbox{-}point({\mathbb S})$).
The last visited node is called \emph{current node} and is
denoted $n$.
The usual notion of \emph{depth} is associated to the search-tree: the
depth is increased by one between a node and its children.  The
function $\delta$ assigns to a node $n$ its depth $\delta(n)$.
Therefore, the state of the search-tree is a quadruple:
$\mathbb{T}=({\cal N}, \Sigma, \delta, n)$.
\end{definition}
In the initial solver state, $n_0$ denotes the root of the search-tree
and all the sets that are part of $\mathbb{S}$
are empty.
\begin{figure}[ht] \footnotesize 
\noindent
\begin{center}
\reglecontroleshort{\newVariable{}}
{<\varset \et {\cal D}>}
{<\varset \cup \{v\} \et {\cal D} \cup \{(v,  {\cal D}_{v,i})\}>}
{\left\{\begin{array}{l}
v \not\in \varset \et \\
{\cal D}(v) = D_{v,i}
\end{array} \right\}}
\saut
\reglecontroleshort{\newConstraint{}}
{<\consset>}
{<\consset \cup \{c\}> }
{\left\{\begin{array}{l}
c \not\in \consset \et  \\
\Var{c} \subseteq \varset
\end{array} \right\}}
\saut
\reglecontroleshort{\post{}}
{<A>}
{<A \cup \{(c,\perp)\}>}
{\left\{\begin{array}{l}
c \in \consset \et  \\
c \not\in \sigma
\end{array} \right\}}
\saut
\reglecontroleshort{\newChild{}}
{<{\cal N} \et \Sigma \et \mathbb{S}>}
{<{\cal N} \cup \{n\} \et \Sigma \cup \{(n,\mathbb{S})\} \et n>}
{\left\{\begin{array}{l}
ch\hbox{-}pt(\mathbb{S}) \et \\
n \notin {\cal N}
\end{array} \right\}} 
\saut
\reglecontroleshort{\jumpTo{}}
{<\mathbb{S} \et \nu>}
{<\Sigma(n) \et n> }
{\left\{\begin{array}{l}
n\not = \nu \et  
n \in {\cal N} \et  \\
ch\hbox{-}pt(\mathbb{S})
\end{array} \right\}} 
\saut
\reglecontroleshort{\solution{}}
{<{\cal N} \et \Sigma \et \mathbb{S}>}
{<{\cal N} \cup \{n\} \et \Sigma \cup \{(n,\mathbb{S})\} \et n> }
{\left\{\begin{array}{l}
sol(\mathbb{S}) \et  \\
 n \notin {\cal N}
\end{array} \right\}} 
\saut
\reglecontroleshort{\failure{}}
{<{\cal N} \et \Sigma \et \mathbb{S}>}
{ <{\cal N} \cup \{n\} \et \Sigma \cup \{(n,\mathbb{S})\} \et n> }
{\left\{\begin{array}{l}
flr(\mathbb{S}) \et \\
 n \notin {\cal N}
\end{array} \right\}} 
\saut
\reglecontroleshort{\deactivate{}}
{<\sigma> }
{<\sigma - \{c\}> }
{\left\{\begin{array}{l}
c \in \sigma
\end{array} \right\}}
\saut
\reglecontroleshort{\restore{}}
{<{\cal D}(v)>}
{<{\cal D}(v) \cup \Delta_v> }
{\left\{\begin{array}{l}
v \in \varset  \et \\
\Delta_v \cap {\cal D}(v) = \emptyset \et 
\Delta_v \subseteq {\cal D}_{v,i}
\end{array} \right\}}
\end{center}
\caption{OS of GenTra4CP (control, without the parameter $\delta$)}
\label{control:MI:figure}
\end{figure}

\begin{figure}[ht] \footnotesize 
\noindent
\def\et{,\ \ }
\begin{center}
\reglecontroleshort{\reduce{}}
{< {\cal D}(v) \et S_e \et A >}
{< {\cal D}'(v)\et S'_e \et A'>}
{\left\{\begin{array}{l}
{\cal D}'(v) = {\cal D}'(v)- \Delta_v^c \et 
\hbox{supprime}\ \Delta_v^c \et  \\
(c, a) \in A  \et  
v \in \Var{c} \et  
\hbox{Red. gén. }\bar{a} \et  \\
A' = A - (c, a) \et  
S'_e = S_e \cup \bar{a}
\end{array} \right\}}
\saut
\reglecontroleshort{\suspend{}}
{< A  \et S_c >}
{< A - \{(c, a)\} \et S_c \cup \{c\} >}
{\{ (c, a) \in A \} }
\saut
\reglecontroleshort{\solved{}}
{< A \et E >}
{< A - \{(c, a)\} \et E \cup \{c\} >}
{\left\{\begin{array}{l}
(c, a) \in A \et \\
solved(c, {\cal D})
\end{array} \right\}}
\saut
\reglecontroleshort{\reject{}}
{< A \et R >}
{< A - \{(c, a)\} \et R \cup \{c\} >}
{\left\{\begin{array}{l}
(c, a) \in A \et \\
false(c, {\cal D})
\end{array} \right\}}
\saut
\reglecontroleshort{\awake{}}
{< A \et S_c >}
{< A \cup \{(c, a)\} \et S_c - \{c\} >}
{\left\{\begin{array}{l}
c \in S_c \et  
a \in S_e \cup \{\perp\} \et \\
awcond(c,a)
\end{array} \right\}}
\saut
\reglecontroleshort{\schedule{}}
{< S_c \et S_e >}
{< S'_c \et S'_e >}
{\left\{\begin{array}{l}
c \in S_c \et  
e \in S_e  \et \\
action(c,a) 
\end{array} \right\}}
\end{center}
\caption{OS of GenTra4CP (propagation)}
\label{propa:events:figure}
\end{figure}
{\tt End of Citation}

The remaining description consists of the description of each event type of the actual trace (called in \cite{ercimlnai} ``generic trace schema'') by introducing other attributes (some of them are redundant like external and internal constraint identifier).
\begin{table}[t] \footnotesize 
\begin{minipage}[t]{0.45\linewidth}
\begin{tabular}{p{0.50\linewidth}p{0.50\linewidth}}
\multicolumn{2}{c}{{\bf Control}}\\
\newVariable{} & $v$,\quad $D_{v,i}$\\[1pt]
\newConstraint{}& $c$\\[1pt]
\post{}, \deactivate{}& $c$\\[1pt]
\restore{}& $v$,\quad $\Delta_v$\\[1pt]
\newChild{}&$n$\\[1pt]
\jumpTo{}& $n$,\quad $n'$\\[1pt]
\solution{}, \failure{}& $n$\\[1pt]
\end{tabular}
\end{minipage}
\hfill\hfill\hfill\vline\hfill
\begin{minipage}[t]{0.45\linewidth}
\begin{tabular}{p{1.5cm}p{3cm}}
\multicolumn{2}{c}{{\bf Propagation}}\\
\reduce{}&$c$,\quad $v$,\quad $\bar{a}$, \\[1pt]
 \quad \quad \quad & \quad $\Delta_v^c$, \quad a \\[1pt]
\suspend{}, \solved{}&$c$\\[1pt]
\reject{}& $c$,\quad $a$\\[1pt]
\awake{}& $c$,\quad $a$\\[1pt]
\schedule{}& $c$,\quad $a$
\end{tabular}
\end{minipage}
\begin{center}
\caption{Attributes of the actual trace of GenTra4CP}
\label{tab:attributes}
\end{center}
\end{table}
 One illustrates the methodology of GT construction by analyzing one ``implementation'' of the GT as presented in \cite{ercimlnai}. In this paper three ``specializations'' of the GT are detailed for three solvers (GNU-Prolog, Choco and PaLM). They consist of a description of the operational semantics of each solver by their transition function.
We show here that the proposed OS for PaLM \cite{jussien00} is compliant. Among the three experimented solvers, PaLM has a clearly different semantics. The transition part of the OS is depicted in the figures~\ref{control:MI:figure} and \ref{propa:events:figure}. 

In order to show that the OS (trace semantics) of PaLM is compliant, one need the following properties of the GT:

(G1) $sol(\mathbb{S}) \Rightarrow R = \emptyset$

(G2) $flr(\mathbb{S}) \Leftrightarrow R \neq \emptyset$

(G3) $(evtype = {\sf reduce}) \Rightarrow R = \emptyset$

(G4) $(evtype = {\sf awake}) \Rightarrow (R = \emptyset \wedge A = \emptyset)$

(G5) $(evtype = {\sf schedule}) \Rightarrow (R = \emptyset \wedge A = \emptyset)$
\begin{figure}  \footnotesize
\reglecontrole{\newVariable{}, \newConstraint{} }{}{}{{\small\text{idem GenTra4CP}}}
\saut
\reglecontrole{\post{} \et \newChild{}}{}{}{{\small\text{idem GenTra4CP}}}
\saut
\reglecontrolespeca{\solution{}}
{<{\cal N} \et \Sigma \et \mathbb{S}>}
{<{\cal N} \cup \{n\} \et \Sigma \cup \{(n,\mathbb{S})\} \et n> }
{\left\{\begin{array}{l}
sol(\mathbb{S}) \et  \\
 n \notin {\cal N} 
\end{array} \right\}} 
\saut
\reglecontrolespeca{\failure{}}
{<{\cal N} \et \Sigma> }
{<{\cal N} \cup \{n\} \et \Sigma \cup \{(n,\mathbb{S})\} \et n> }
{\left\{\begin{array}{l}
 n \notin {\cal N} \et  \\
R \not = \emptyset 
\end{array} \right\}} 
\saut
\reglecontrole{\deactivate{}}{}{}{{\small\text{idem GenTra4CP}}}
\saut
\reglecontrolespeca{\restore{}}
{<{\cal D}(v) \et Q_t \et {\cal E}>}
{<{\cal D}(v) \cup R_v \et Q_t \cup \bar{a} \et {\cal E} - E>}
{\left\{\begin{array}{l}
v \in {\cal V} \et  
R_v \subseteq \{d \in \mathbb{D} | {\cal E}(v,d) \cap \sigma \neq \emptyset\} \et  \\
E = \{{\cal E}(v,d) | d \in R_v\} \et \\
\bar{a}\ \hbox{actions de restauration de}\ {\cal D}(v)
\end{array} \right\}}
\caption{OS of PaLM \cite{ercimlnai} (control)}
\label{fig:rules:palm:control}
\end{figure}

\begin{figure}[ht]
\reglecontroleshort{\reduce{}}
{<{\cal D}(v) \et Q_t \et {\cal E}> }
{\left\{\begin{array}{l}
{<\cal D}(v) - \Delta_v^{c_{a}} \et Q_t \cup \{\bar{a}\} \et \\
{\cal E} \cup \{(v,d,C) \,|\, d \in \Delta_v^{c_{a}}\}>
\end{array} \right\}}
{\left\{\begin{array}{l}
 v \in \Var{c} \et 
  R=\emptyset  \et 
 A=\{(c,a)\}  \et \\
 \Delta_v^{c_{a}} \neq \emptyset \hbox{ set of inconsistent values for $v$} \et \\
 C \subseteq \sigma \hbox{ explains the removal of $\Delta_v^{c_{a}}$ from ${\cal D}(v)$} \et \\
 \hbox{The reduction generates } \bar{a}
 \end{array}\right\} }
\saut
\reglecontroleshort{\suspend{}}
{< A  \et S_c >}
{< \emptyset \et S_c \cup \{c\} >}
{\{ A = \{(c, a)\} \} }
\saut
\reglecontroleshort{\reject{}}
{< A \et R >}
{< \emptyset \et R \cup \{c\} >}
{\left\{\begin{array}{l}
A = \{(c, a)\} \et 
v \in \Var{c} \et 
{\cal D}(v)=\emptyset
\end{array} \right\}}
\saut
\reglecontroleshort{\awake{}}
{<S_c \et A> }
{<S_c - \{c\} \et \{(c, a)\}>}
{\left\{\begin{array}{l}
A = \emptyset \et 
c \in S_c  \et 
R=\emptyset  \et \\
a \in Q_h \cup \{\perp\} \et 
dependence(c,a)
\end{array}\right\}}
\saut
\reglecontroleshort{\schedule{}}
{<Q_h \et Q_t> }
{<\{a\} \et Q_t - \{a\}>}
{\left\{\begin{array}{l}
select(a) \et 
A = \emptyset \et 
a \in Q_t \et 
R = \emptyset 
S_c \not = \emptyset
\end{array}\right\}}
\saut
\caption{OS of PaLM \cite{ercimlnai} (propagation)}
\label{fig:rules:palm:propag}
\end{figure}
One admits:

(P1) $dependence(c,a) \Leftrightarrow  awcond(c,a)$

(P2) $select(a) \Rightarrow \exists c \in {\cal C} \ action(c,a)$

(P3) $\exists v \in \Var(c), {\cal D}(v) = \emptyset \Rightarrow false(c, {\cal D})$

\begin{theorem}
\label{prop:proofpalmderiv}
\ \ 

\vspace{1mm}
The GT restricted to all events depicted in the Figures~\ref{control:MI:figure} and \ref{propa:events:figure} but \jumpTo{} and \solved{}, is a parametric subtrace of GenTra4CP, derived from the trace specified for PaLM (Figures~\ref{fig:rules:palm:control} and \ref{fig:rules:palm:propag}). 
\end{theorem}
%

\section{Generic Trace and Constraints Specification}
\label{tragenplc}

This approach of semantics can be applied to constraints specification. The question then is whether it exists a generic trace covering all the constraints that one wishes to describe, i.e. covering different types of constraints (single, global, ...), different domains ( FD, intervals, ...), different classes of solvers (CSP, SAT, rules, such as CHR), different levels (algorithms, modules, modeling) or different aspects (language, interaction, interfaces, ...) as well.

We limit ourselves here to the CSP case. Each constraint has a declarative semantics defined by the relation it represents on its domains. The GT can thus provide a description of the possible effects of each constraint separately or in a network, regardless the particular algorithm it implements. In this sense such semantics is a kind of minimal description of what we should be able to observe of the behavior of a constraints set. It can be used to define any kind of interfaces, particularly for problem modeling.

In practice, as is what has been done for GenTra4CP, one should start with a definition of an actual trace whose meaning can be given by a reconstruction function. It should be completed by an OS as large as possible such that parameters relevant to potential interfaces and applications are fully described.

\vspace{1mm}
We illustrate this approach of a generic semantics with a simple resolution example, showing the two traces obtained with GNU-Prolog and PaLM for this example. Both solvers have been instrumented to produce the generic trace for CSP(FD), and their traces can be ``understood'' using the OS of the Figure~\ref{fig:gentra4cpexa}.
\begin{figure}[ht] \footnotesize 
\noindent
\begin{center}
\reglecontroleshort{\newVariable{}}
{[\newVariable{},\quad v,\quad D_{v,i}]}
{
<\varset \et {\cal D}> \rightarrow 
<\varset \cup \{v\} \et {\cal D} \cup \{(v,  D_{v,i})\}>	
}
{\left\{\begin{array}{l}  
\end{array} \right\}}
\saut
\reglecontroleshort{\newConstraint{}}
{[\newConstraint{},\quad c]}
{
<\consset> \rightarrow 
<\consset \cup \{c\}> 
}
{\left\{\begin{array}{l}  
\end{array} \right\}}
\saut
\reglecontroleshort{\post{}}
{[\post{},\quad c]}
{
<{A}> \rightarrow 
<A \cup \{(c,\perp)\}>	
}
{\left\{\begin{array}{l}  
\end{array} \right\}}
\saut
\reglecontroleshort{\newChild{}}
{[\newChild{},\quad n]}
{
<{\cal N} \et \Sigma \et \mathbb{S}>  \rightarrow 
<{\cal N} \cup \{n\} \et \Sigma \cup \{(n,\mathbb{S})\} \et n> 
}
{\left\{\begin{array}{l}  
\end{array} \right\}}
\saut
\reglecontroleshort{\reduce{}}
{[\reduce{},\quad c,\quad v,\quad \bar{a},\quad \Delta_v^c,\quad a]}
{
< {\cal D}(v) \et S_e \et A > \rightarrow 
< {\cal D}(v) - \Delta_v^c \et S_e \cup \bar{a} \et A - (c, a)>	
}
{\left\{\begin{array}{l}  
\end{array} \right\}}
\saut
\reglecontroleshort{\suspend{}}
{[\suspend{},\quad c,\quad a]}
{
< A  \et S_c > \rightarrow 
< A - \{(c, a)\} \et S_c \cup \{c\} >	
}
{\left\{\begin{array}{l}  
\end{array} \right\}}
\saut
\reglecontroleshort{\awake{}}
{[\awake{},\quad c,\quad a]}
{
< A \et S_c > \rightarrow 
< A \cup \{(c, a)\} \et S_c - \{c\} >	
}
{\left\{\begin{array}{l}  
\end{array} \right\}}
\end{center}
\caption{OS of GenTra4CP (reconstruction)}
\label{fig:gentra4cpexa}
\end{figure}
Both traces (Figure~\ref{fig:trace:exampleGNUPalLM}) correspond to the resolution of (GNU-Prolog syntax) {\tt fd\_element\_var(I,[2,5,7],A), (A\#=I ; A\#=2)} which admits one solution only\footnote{PaLM produces shortcuts such that the sequence  \reduce{} \suspend{} \schedule{} \awake{} is displayed as \reduce{} \awake{}. Such shortcut does not have any semantics in GenTra4CP (it could be adapted). This shows only that the PaLM OS given in \cite{ercimlnai} was not actually compliant to the GT.}. The declarative semantics of this constraint (all variables are finite domain) can be expressed as: 
{\tt fd\_element\_var(I, L, V)} ({\tt L} liste) constrains {\tt V} to be equal to the {\tt I}ith element of {\tt L}. Thas is to say all triples such that $i \in$ {\tt I}, $u \in$ {\tt L}($i$), $v \in$ {\tt V} and $u=v$ are valid. The interval {\tt [}$a$-$b${\tt ]} denotes \emph{from $a$ to $b$} and {\tt [}$a$,$b${\tt ]}, \emph{$a$ and $b$}.
\begin{figure}
\begin{scriptsize}
\begin{minipage}[t]{0.47\linewidth} \scriptsize 
\begin{verbatim}
 1[0]choice point node(0)
 2[1]newVariable v1 [0-mx]
 3[1]newVariable v2 [0-mx]
 4[1]newConstraint c1
fd_element([v1,[2,5,7],v2])
 5[1]post c1
 6[1]reduce c1 v1 [0,4-mx]
 7[1]reduce c1 v2
   [0-1,3-4,6,8-mx]
 8[1]suspend c1
 9[1]choice point node(1)
10[2]newConstraint c4
    x_eq_y([v2,v1])
11[2]post c4
12[2]reduce c4 v2 [5,7]
13[2]reduce c4 v1 [1,3]
14[2]suspend c4
15[2]schedule v2 dom
16[2]awake c1
17[2]reject c1
18[2]failure node(2)
...
\end{verbatim}
\end{minipage}
\hfill\hfill\vline\hfill
\begin{minipage}[t]{0.47\linewidth} \scriptsize 
\begin{alltt}

 0[0]newVariable v0 I [0-mx]
 1[0]newVariable v1 A [0-mx]
 2[0]newConstraint c0 
     element(I,[2,5,7],A)
 3[0]post c0
 4[0]suspend c0
 5[0]awake c0            
 6[0]reduce c0 v0 [3-mx] max
 7[0]reduce c0 v1 [0,1] min
 8[0]reduce c0 v1 [8-mx] max
 9[0]suspend c0
10[0]newConstraint c1 eq(I,A)
11[0]post c1
12[0]suspend c1
13[0]awake c0 (v0,max)
14[0]reduce c0 v1 [2-7] empty
15[0]reject c0 empty
16[0]failure
17[0]newVariable v-1 I [0-1]
18[0]reduce c2 v-1 [0,1] empty
...
\end{alltt}
\end{minipage}
\end{scriptsize}
\caption{Partial actual trace of GNU-Prolog and PaLM with the given example. The second attribute is the choice-tree depth}
\label{fig:trace:exampleGNUPalLM}
\end{figure}
One may observe\footnote{GenTra4CP produces traces in  XML, readable but verbose. A more concise representation has been adopted here.} that the traces are different, so, in particular:
\begin{itemize}
\item the domain of {\tt I} is not the same for GNU ({\tt [1-3]}) and for PaLM ({\tt [0-2]});
\item the order and the values of the values removal are not the same, as the choice of variables to consider;
\item search spaces are different;
\item a specific variable occurs in the trace of PaLM ({\tt v-1}).
\end{itemize}
These variations are irrelevant when comparing the respective semantics (renaming, extra variable) and from both actual traces one may reconstruct the corresponding virtual ones. However some variations should be examined and fixed like the limit values of {\tt I}, or some specific attributes.

\section{Discussion}
\label{discussion}

The semantics of traces presented here corresponds to the ``Observable Semantics'' of Lucas \cite{lucas00} or the partial trace semantics of Cousot \cite{CousotPOPL02}. The parameters of the virtual states are, as expressed by Lucas, ``syntactic objects used to represent the conduct of operational mechanisms''. The traces are abstract representations of process semantics which allow to take into account the sole details we want to consider as common to a set of processes.
The choice to relate two forms of trace (virtual and actual) corresponds to the need to reconcile different pragmatic approaches: formal specification of semantics more or less abstract, and empirical manipulations of traces like in trace-based systems \cite{Lotfithese11e}. We established here a particular method to demonstrate compliance of a process trace with regards to a generic trace. This approach allows to establish formal relations with the trace theory \cite{dieroz95} too.

\vspace {1mm}
We have shown here that the definition of the trace GenTra4CP can be well defined in such a theoretical framework, and we have characterized by relatively simple transformations (parametric subtrace, similarity and derivation) the formal linkages between the observed processes and the generic trace. This analysis revealed some insufficiencies in the formal definition of GenTra4CP as the lack of formal verification of particular traces solver compliance. 
Simonis \& al \cite{simonisDFMQC10} note that the generic trace GenTra4CP contains too many details with a too sophisticated specification. This is certainly true if the objective is just to analyze the evolution of some problem variables and some aspects of the search. In this case the need of trace information is limited and it is less work to implement directly the capture of the needed information rather than implementing a full generic trace format. But it is different if the objective is to create a generic interface between solvers and many more applications. Our study shows also that GenTra4CP probably contains too many optional details with no clear semantics, such that implementers feel free not to implement many of them, or to implement them with just specific implementation dependent semantics. A more demanding approach, but which may be more useful, could be to specify formally more attributes of the generic trace.

Moreover, as it has been observed in the Section~\ref{sec:tragen}, it is the task of the developer of a solver to implement a generic (sub)trace or to adapt the tools which have been developed on the basis of the generic trace. The investment to make is measured by the gap between the developed process trace and the generic trace (formally a derivation, Figure~\ref{fig:utilgen}). It may seem easier to implement an ad-hoc trace systematically, rather than to implement once a compliant tracer, or to adapt a tool each time needed. Langevine and Ducass\'e have shown \cite{langevine08} that a generic approach could have more advantages than drawbacks, but it is similar to a standardization effort.

\vspace{1mm}
Such an effort can only result from the action of a large community, and not from a small group as in the case of GenTra4CP. The project of standard \cite{standcp10} focuses mainly on the definition of a Java interface that includes in particular the major types of variables, unary constraints, some binary and global constraints, as some strategies to search for solutions. But the question of the semantics cannot be ignored. If the declarative semantics of simple constraints poses little problem of specification, it is not the same for the operational semantics, whose accuracy depends on potential applications developed with constraint problems. The approach presented here, based on a generic trace semantics, may be a way since it provides a framework for specifying outcomes and side effects of constraint, revealing for example constraints interactions independently from specific implementations.

\section{Conclusion}
\label{conclusion}

GenTra4CP has been an innovative approach using a partial trace semantics to handle both problems of specifying constraint solvers (on finite domains) and of portable analysis tools. Such an effort was similar to a standardization effort, but with no effective dissemination because of its limits (small group who made it, some technical gaps and restricted to one constraints domain).

We have introduced a simple formal framework based on trace theory and abstract interpretation to explain the method of generic trace construction, and to show the potential value of this approach to specify a partial semantics of constraints resolution.

The realization of a generic trace for a significant set of simple or global constraints certainly represents a considerable amount of efforts. It seems however that such an approach could not only allow the portability of potential applications, but also contribute to the semantics of knowledge representation systems which combine several methods like constraints and rules.



\bibliographystyle{splncs03}
\bibliography{LTS,oadymppac}

\clearpage
\section*{ANNEX: Proofs}
\label{appendixAD}

\subsection{Proof of theorem~\ref{prop:proofpalmderiv}}
\label{proof:proofpalmder}

One shows that a parametric subtrace of PaLM is simulable by a parametric subtrace of GenTra4CP.

\vspace{1mm}
One considers the GT GenTra4CP, restricted to all the events of the Figures~\ref{control:MI:figure} and \ref{propa:events:figure} but \jumpTo{} and \solved{}. Ignoring \jumpTo{} does not affect the search-tree construction but only its visiting strategy, and ignoring  \solved{} corresponds to removing the parameter $E$ in the solver state. Furthermore the parameter corresponding to the explanations can be ignored, as it is not formalized in the OS of GenTra4CP. 

According to the definition~\ref{def:stacceptable}, the restriction to the subset of considered events is a parametric subtrace of GenTra4CP.

The subtrace of PaLM to be considered consists just in ignoring the explanations. This does not restrict the set of action types (definition~\ref{def:stacceptable}).

\vspace{1mm}
In GenTra4CP and the considered subtrace, the control part $\mathbb{T}_g$ uses actually 4 parameters:\quad
${\cal N}, \Sigma, \delta, \nu$, and the propagation part $\mathbb{S}_g$ 8 parameters:\quad
${\cal V}, {\cal C}, {\cal D}, A, E, R, S_c, S_e$, in total 12 parameters.

In PaLM, the control part $\mathbb{T}_p$ uses 5 parameters:\quad
${\cal N}, \Sigma, \delta, \nu, Q_t$, and the propagation part $\mathbb{S}_p$ 9 parameters:\quad
${\cal V}, {\cal C}, {\cal D}, A, R, S_c, Q_h, Q_t, {\cal E}$; in total 13 parameters ($Q_t$ is common).
There are some differences:
\begin{itemize}
\item $E$, the subset of the ``constraint store'', containing the valid constraints, is irrelevant in PaLM, as no satisfiability test is realized in PaLM (no ``entailment'').
\item The set $S_e$ of the current events in PaLM is a queue ($S_e = Q_h \cup Q_t$) whose head $Q_h$ (a singleton) contains the selected current event.
\item The state of the PaLM solver contains an additional parameter ${\cal E}$, the explanation function which serves to store the what is called the ``explanations''. $E$ is a partial function: ${\cal E}: {\cal V} \times \mathbb{D}
\longrightarrow {\cal P}(\sigma)$\footnote{${\cal P}(\sigma)$: powerset of the store $\sigma$ (instance of the constraints which are in $A$, $S_c$ and $R$ for PaLM).} which assigns to each value removal $(v, d)$ ($v \in \varset, d \in {\cal D}(v)$) a set of non relaxed constraints which explains this removal. This partial function is updated by the events \reduce{} and \restore{}.
\item $A$, in PaLM, has at most one element.
\end{itemize}

\vspace{1mm}
Thus one shows that the PaLM subtrace is simulable in the subtrace ``PaLM'' of GenTra4CP. 

One uses the theorem~\ref{prop:proofmethabstr}. One defines the application $d$ between the modified states $\mathbb{T}_p \times \mathbb{S}_p$ and $\mathbb{T}_g \times \mathbb{S}_g$, that is: (one omits $\delta$ which is deducible directly from ${\cal N}$)

${\cal N}, \Sigma, \nu, {\cal V}, {\cal C}, {\cal D}, A, R, S_c, Q_h, Q_t$ and

${\cal N}, \Sigma, \nu, {\cal V}, {\cal C}, {\cal D}, A, R, S_c, S_e$

as follows: identity for the 9 first parameters of PaLM ${\cal N}, \Sigma, \nu, {\cal V}, {\cal C}, {\cal D}, A, R, S_c$, then $Q_h \cup Q_t = S_e$. 

\vspace{1mm}
The action types have the same names ans their set is restricted to those in the Figures~\ref{fig:rules:palm:control} and \ref{fig:rules:palm:propag}.

\vspace{1mm}
The initial states $\mathbb{T}_{0,p} \times \mathbb{S}_{0,p}$ and $\mathbb{T}_{0,g} \times \mathbb{S}_{0,g}$ to be considered are:

\noindent
$\{rc_p\}, (rc_p, \mathbb{S}_{0,p}), rc_p, \emptyset_p, \emptyset_p, \emptyset_p, \emptyset_p, \emptyset_p, \emptyset_p, \emptyset_p, \emptyset_p$ and

\noindent
$\{rc_g\}, (rc_g, \mathbb{S}_{0,p}), rc_g, \emptyset_g, \emptyset_g, \emptyset_g, \emptyset_g, \emptyset_g, \emptyset_g, \emptyset_g, \emptyset_g$

\vspace{1mm}
\newVariable{}, \newConstraint{}, \post{}, \newChild{} and \deactivate{} are in correspondence as the transition rules are the same, as their modified parameters as well.

For \solution{} and \failure{}, it is the same provided the properties (G1) and (G2) hold.

The case of \restore{} is more complex. But, if one ignores the explanations and take for $\Delta_v$, $R_v$ ($\Delta_v = R_v$) for the same variable $v$, the conditions associated to the event of  Gentra4CP are deducible from the explanations properties (restitution of the removed values, then inexistent in the current domain of $v$). But one has to justify the update of  $S_e$ in the transition rule of GenTra4CP.

\vspace{1mm}
\reduce{}. To $\Delta_v^{c_{a}}$ it corresponds $\Delta_v^{c}$ (set of inconsistent values) of the GT. By (G3) the properties $R = \emptyset$ correspond. Finally as $d(Q_h \cup Q_t) = S_e$, then $d(Q_h \cup Q_t \cup \bar{a}) = S_e \cup \bar{a}$.

\suspend{}. In the corresponding initial states $(c, a) \in A$, and $A' = A - \{(c, a)\}$ in the final states.

\reject{}. Uses (P3) for the initial states, and the final states are in correspondance.

\awake{}. Uses (P1) and (G4).

\schedule{}. Uses (P2) and (G5). $S_c$ and $S_e$ are invariants in the GT.


\end{document}